\documentclass{sig-alternate-05-2015}
\usepackage{multirow}
\usepackage[ruled]{algorithm2e}
\usepackage{balance}
\usepackage{underscore}

\begin{document}




\conferenceinfo{HIP3ES '16}{January 18--20, 2016, Prague, Czech Republic}


%

\title{Taxim: A Toolchain for Automated and Configurable Simulation for Embedded Multiprocessor Design}
%
%
%
%
%

\numberofauthors{3} 
%
\author{
%
%
\alignauthor
Gorker Alp Malazgirt\\
       \affaddr{Department of Computer Engineering}\\
       \affaddr{Bogazici University}\\
       \affaddr{34342 Bebek, Istanbul, Turkey}\\
       \email{alp.malazgirt@boun.edu.tr}
\alignauthor
Deniz Candas\\
       \affaddr{Istanbuler Gymnasium}\\
       \affaddr{34112 Fatih, Istanbul, Turkey}\\
       \email{dnzcandas@gmail.com}
\alignauthor Arda Yurdakul\\
       \affaddr{Department of Computer Engineering}\\
       \affaddr{Bogazici University}\\
       \affaddr{34342 Bebek, Istanbul, Turkey}\\
       \email{yurdakul@boun.edu.tr}
\and  
}

\maketitle
\begin{abstract}
Multicore embedded systems have been constantly researched to improve the efficiency by changing certain metrics, such as processor, memory, cache hierarchies and their cache configurations. Using Multi2Sim and McPAT simulators in combination allows the user to design various multiprocessing architectures and estimate performance, power, area and timing metrics. However, the design time required to simulate these systems is daunting and prone to human error.  In this paper, we introduce Taxim, a toolchain that can automatically create requested multicore on-chip topologies along with minimizing the simulation time due to repetitive tasks between architectural power, energy and timing simulations. Taxim's decision-tree-based topology synthesis tool creates processor configuration files that can be highly erroneous when generated manually. The toolchain also automates the steps from design entry to output report extraction by running automation scripts, and listing the results. Our experiments show that multiprocessing architectures with 32 cores and irregular cache hierarchies are more than 1k lines of code in Multi2Sim's processor configuration format and Taxim can create such a file in less than 10 milliseconds. The source code is freely available at \textit{https://github.com/bouncaslab/TaXim/}.
\end{abstract}

%
%


%
%

%
%



\section{Introduction}
The current technological challenge faced with embedded systems is balancing for varying domains performance, energy efficiency, and production costs that are dominated by the design time and the chip area. The simulation tools are essential for designing complex systems because these tools help the designer in searching the design space and without experimenting on physical implementations, which are time consuming and costly. In the embedded systems design domain, cache \cite{Tarjan2006}, memory \cite{Rosenfeld2011}, architectural \cite{Ubal2007},\cite{Binkert2011} and energy \cite{Li2009} simulators exist so as to evaluate the energy or performance of processor architectures in the large design exploration space. Thus, simulators are coupled with design space exploration (DSE) tools \cite{Zaccaria2010},\cite{Desmet2010} where optimal designs are sought using algorithmic approaches based on designer's requirements. In general, this procedure consists of a large number of iterations between DSE's optimizer and the simulators, and continues until sufficiently good results are achieved. Hence, in general, at each iteration the design under simulation undergoes hardware or software modifications and this process requires automation.

\begin{figure}[t]
\centering
\includegraphics[width=8cm]{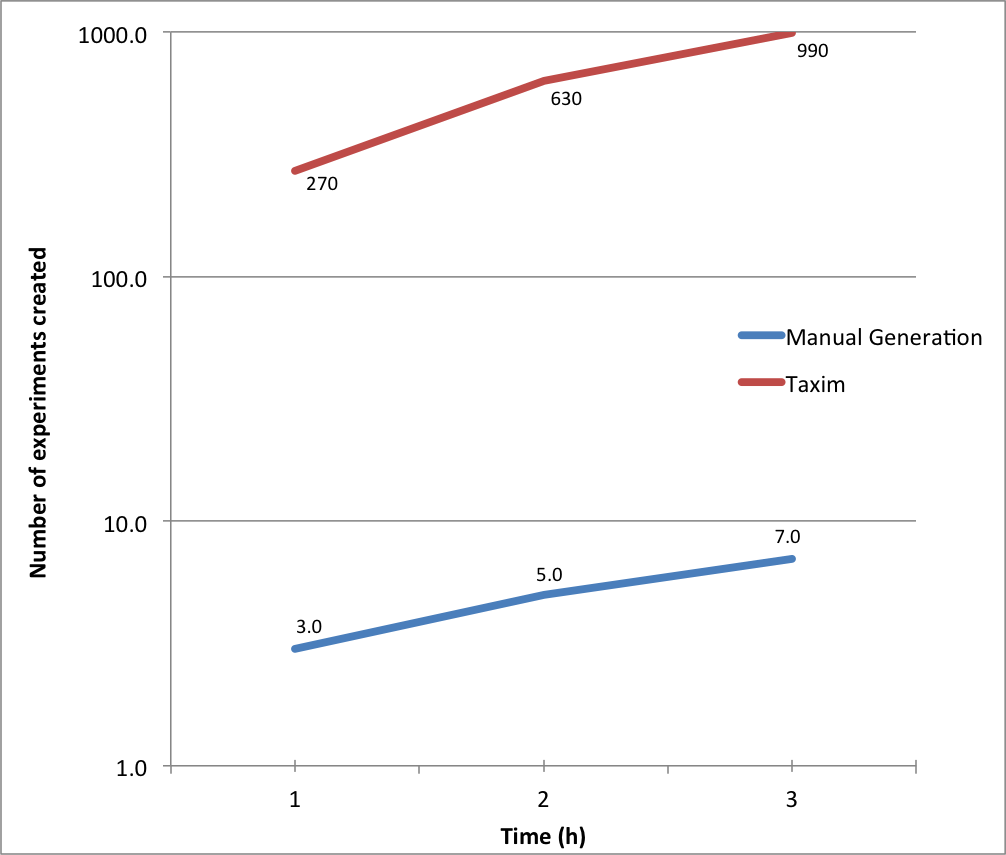} 
\caption{Comparison of manual and automatic experiment generation with respect to time, creating 4-core multiprocessor topologies}
\label{fig:generation}
\end{figure}

Multiprocessor design exploration requires generation of various multiprocessor topologies. These topologies do not only differ in cache organizations and number of cores, but they also vary in the number of caches, cache levels, and data/instruction cache sharing metrics. Specifically, for multiprocessing with many cores and caches, manual cache-processor-memory connectivity creation is time consuming and prone to human error.

In this work we present Taxim, a toolchain that automates architecture generation and the simulation of embedded multiprocessing systems using Multi2Sim \cite{Ubal2007} and McPat \cite{Li2009} simulators that have been used extensively in research and DSE tools. Multi2Sim offers the designer a fully customizable environment that simulates both the devices and their interactions. McPat on the other hand allows a detailed exploration of crucial efficiency metrics.

Automated architecture generation is fast and not prone to factors such as fatigue. In Figure \ref{fig:generation}, we present the number of experiment creation with respect to time. Experiment creation task includes the generation of 4-core multiprocessor topology lists that have less than 200 lines of Multi2Sims \cite{Zaccaria2010} architecture representation files, automation scripts and simulation output preparation. Based on our observations on a group of three students, we have measured that the rate of design generation decreases over time due to fatigue. Thus, However, with an automated tool, after an initial setup time to specify design metrics and prepare a topology list, experiment preparation continues at a constant pace.

In this paper, Taxim's contributions can be summarized as:
\begin{enumerate}
\item In Taxim, we introduce a decision tree based method for that creates processor multi-core processing architectures by using decision trees. Thus, Taxim speeds up design time considerably when compared with the manual generation of the architectures. 
\item It provides an automated toolchain that allows designers to run multiple simulations using Multi2Sim \cite{Ubal2007} and McPat \cite{Li2009}. In addition, Taxim is built to prepare structured output reports of numerous results such as processor datapath specifics, cache line utilization to leakage power consumption. Design space exploration (DSE) tools \cite{Zaccaria2010,Desmet2010} can utilize these reports. 
\item While setting up the environment for multiple simulations, the designer might enter some parameters wrong. Hence, the simulations might have to get aborted. Yet, the existence of erroneous simulation set-ups might truly slow down the completion time of multiple simulations. In order to prevent erroneous simulations due to human factor, we introduce intelligent control in Taxim. In this way, multiple simulations can be carried out smoothly in the shortest possible time.   
\end{enumerate}

We present Taxim in the following way. We first present related works in the next section. Section 3 details our architecture generation and validation method. In Section 4 we detail Taxim's process flow. Section 5 presents the experimental results that compare Taxim and manual approaches. We conclude the paper in Section 6 with our conclusions.
 \begin{figure}[t]
\centering
\includegraphics[width=8cm]{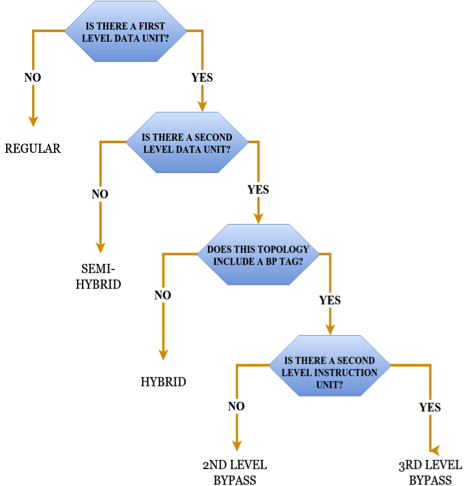} 
\caption{Decision tree that represents different pathways taken by Taxim to create a given topology}
\label{fig:tree}
\end{figure}

\section{Related Work}
Simulators have been an important part of designing processing systems \cite{Binkert2011, Yan2015,Chis2013,Mohanty2013,Thompson2013,Devigo2015}. The first step before running a simulation is design entry, that allow designers to customize the hardware and the benchmark that will execute. Our work differs from available works as follows:

First, we present an automated processor architecture generation method based on a decision tree. It allows us to generate very complicated multi core architectures with multiple levels of cache hierarchy with ring bus network support when necessary. Automatic architecture generation tool also validates cardinality and connectivity of generated architectures. This allows automatic generation more advantageous over manual generation specifically for complex designs. 

Second, we present an automatic toolchain, which automates the step from design entry to output, report extraction for architectural and power, area and timing simulations. In this work, unlike existing research works we present these steps decoupled from other topics such as design space exploration \cite{Yessin2014,Ventroux2010,Javaid2014} or simulation methods \cite{Ventroux2010}. In this way, our tool-kit can be used as a plug-in by the design space exploration tools. Since the environment enables multiple concurrent simulations, the simulation time can be considerably reduced in multi-core machines and supercomputers. Obviously this will reduce the design exploration time when Taxim is coupled with DSE tools.

In the design space exploration literature, there have been plenty of studies that require large number of simulations for finding pareto set or the single "best" design point from single or multiple design parameters \cite{Chis2013,Thompson2013, Ventroux2010,Javaid2014,Bengueddach2013,Jia2010,Yessin2014,Carlson2011,Devigo2015}. For this reason, this work exhibits an automatic architecture topology generation method that can be used in DSE systems that employ Multi2Sim \cite{Zaccaria2010} and McPat \cite{Calborean2011}. 
Alongside automated architecture generation, automated design entry to extraction of output reports allows DSE tools to process design points in a rapid and continuous manner. The authors in \cite{Bengueddach2013} explore energy consumptions of various two level cache configurations using McPat simulations. There has not been any automation regarding the architectural simulations due to the fact that the underlying processor architecture is fixed. Similarly, the work in \cite{Jia2010} explores cache hierarchy by estimations and simulations. However, cache topology has been fixed, thus automated cache topology generation has been omitted. However, our work first presents automatic architecture generation, then explains the rest of the tasks that automate whole simulation process from design entry to output extraction, using Multi2Sim and McPat environments.

Authors in \cite{Chis2013} have presented a design space exploration of computer architectures that searches in a large design space. The automated tool customizes cache architectures and processor internals, however there has not been any architectural exploration such as connectivity of different cache levels, cache sharing and processors in Multi2Sim and McPat. In addition, the generation scripts have only been limited to single-core architectures. The work in \cite{Devigo2015} takes a range of high and low-level parameters to improve accuracy in the design of a multiprocessor system on a chip. Authors automate generation of processor specifics such as the instruction set, number of pipelines and pipeline stages. However, cache topology is fixed whereas our architecture generation tool also automates the cache topology creation. 

Cache hierarchy exploration engine presented by Yessin et al \cite{Yessin2014} first determines the boundaries of the design space by determining the ranges of the design variables and prepares the memory traces through simulation. Next, an optimization algorithm iteratively determines the most suited cache hierarchy from the given benchmark set. By using an automated toolchain similar to Taxim, the optimization unit of the system can request additional simulations, thus the feedback loop can be expanded with more simulation results. Similarly, the work in \cite{Ventroux2010} provides an mpsoc simulation environment where designers can provide automation scripts to generate simulations with various design parameters via the parameter file. The presented environment allows embedding different memory units with different Network-On-Chip (NOC) architecture support. In this way, the simulation tool can be connected to various design automation tools. The design space can be enlarged by combining simulation and estiomation of architectures with tolerable error margins. The authors in \cite{Javaid2014} also provides a method for performance estimation of pipelined multiprocessor system-on-chip architectures. Analytical models are combined with simulation data and exploration is handled on the aggregated design space. The design of the experiments requires numerous experiments therefore both simulation and estimation requires architectural preparation and application settings. Thus, when the number of architectures and benchmarks increases, a toolchain like Taxim plays an important role for automating architectural preparation, simulation/estimation and results extraction tasks.

\begin{algorithm}
 \KwIn{An architecture's string representation}
 \KwOut{Multi2Sim processor configuration file}
 {\bf Step 1:} Parse given architectural string and tokenize cores, instruction caches and data caches;\\
 {\bf Step 2:} Determine number of cores, then determine the number of  first, second and third level instruction and data caches, group them with respect to levels;\\
 {\bf Step 3:} Traverse the decision tree and select architecture type from available types: (Regular, Semi-Hybrid, Hybrid, 2nd Level Bypass and 3rd Level Bypass);\\
 {\bf Step 4:} For each core in the system:\\ 
Connect each core to an empty first level data cache;\\
If there does not exist an empty  first level data cache;\\
Connect each core to the least connected first level data cache;\\
Connect each core to an empty first level instruction cache;\\
If there does not exist an empty first level data cache;\\
Connect each core to the least connected first level instruction cache;\\
{\bf Step 5:} For each level in the cache hierarchy:\\
Connect each data/instruction cache to an empty upper level data cache\\
If there is no empty cache;\\
Connect each core to the least connected upper level data cache;\\ 
If there is no upper level cache;\\
Connect each cache to the memory;\\
{\bf Step 6:} If BP exists in the string representation:\\
    For each last level memory component connect a network switch:\\
    For each incoming cache connection to each: last level memory \\
    For each cache level in the system do separately for each data and instruction cache:\\
    Group data/instruction cache connections in two and connect to a switch;\\
    Connect each switch pair with a new switch;\\
    If total number of caches is odd, connect the remaining cache to one of the pairing cache switch;\\
    Connect each switch pairs to the memory;\\ 
 \caption{Architecture generation from string representation}
 \label{alg:topologygen}
\end{algorithm}  

\begin{table}
\centering
\begin{tabular}{|c|c|}
\hline 
\textbf{Symbol} & \textbf{Definition}\tabularnewline
\hline 
\hline 
C & Core Count\tabularnewline
\hline 
\multirow{2}{*}{L$<$x$>$} & Shared Data/Instruction \tabularnewline
 & Level 1, 2 or 3 Cache\tabularnewline
\hline 
IL$<$x$>$ & Instruction Level 1 or 2 Cache\tabularnewline
\hline 
DL$<$x$>$ & Data Level 1 or 2 Cache\tabularnewline
\hline 
BP & Exists a bypass connection\tabularnewline
\hline 
\end{tabular}

\caption{Nomenclature of the Topology Representation}
\label{tbl:nomen}
\end{table}

\section{Details of Architecture Generation}
In this section we detail the architecture generation and validation methods. Our nomenclature is presented in Table \ref{tbl:nomen}. We define the number of cores with C. When data and instruction caches are shared, we represent these caches as L. When instruction and data caches are separate, they are specified as IL and DL respectively. If there exists a bypass connection \cite{Malazgirt2015}, it is shown with BP. We can show usage of the nomenclature in two examples:
\begin{itemize}
\item 2C\_4L1\_1L2 explains that there exists two cores, four L1 shared caches and one L2 shared cache
\item 2C\_2DL1\_2IL1\_1DL2\_BP shows us that there exists two cores with two data L1 cache, two instruction L1 caches, one data L2 cache and there exists a bypass connection from instruction caches to the memory
\end{itemize}

This nomenclature is followed to create topology representations of designated architectures. The designer can enter in string format to receive configurations recognizable by the simulator. The designer can also opt to work with provided topology sets that are available in the library of the toolchain. The topology creation tool relies on strictly defined "Topology Creation Rules" as outlined in \cite{Malazgirt2015}. The user may specify the desired topologies in a text file by entering their names according to the designated nomenclature.

\textbf{Architecture Generation}: During architecture generation, Taxim first tries to extract the type of topology by parsing the names of the architectures. The parser also checks compliance of the architecture name to the  "Topology Generation Rules" \cite{Malazgirt2015}. Common design rule errors are missing underscores, wrong order of caches, missing or extra letters that do not conform to generation rules. After parsing the nomenclature without any errors, a simple decision tree is employed as shown in Figure \ref{fig:tree}. The decision tree (DT) guides the generation tool for the type of routing that should be applied to construct the given topology.

The decision tree consists of four rules to call five different methods. These methods are: Regular Fat Tree, Semi-Hybrid, Hybrid, 2nd Level Bypass and 3rd Level Bypass. The generation rules can be summarized as follows and they are derived from "Topology Creation Rules" \cite{Malazgirt2015}:
\begin{itemize}
\item Rule 1: Taxim checks if there exists separate data and instruction caches. If not, the generator employs Regular method where Von Neumann or Harvard architectures can be employed
\item Rule 2: Taxim checks if there exists second level data unit. Absence of second level data caches employs Semi-Hybrid method. In this method, the generated architectures are hybrid architectures because there exists shared instruction caches at the first level. Nevertheless, above first level, all data and instruction caches are shared without any bypassing
\item Rule 3: Taxim checks if given topology representation has BP tag that stands for by-passing a cache. Absence of BP tag means that the given architecture is a hybrid architecture and data caches are separate from instruction caches at all levels
\item Rule 4: Taxim checks if there is a second level instruction cache. The absence of second level data cache employs the 2nd level bypass method that generates architectures with first level instruction caches bypassing second level cache to main memory. Finally, the 3rd level bypass method is employed that means second level instruction caches are bypassed third level caches.     
\end{itemize}

\begin{figure*}
\centering
\includegraphics{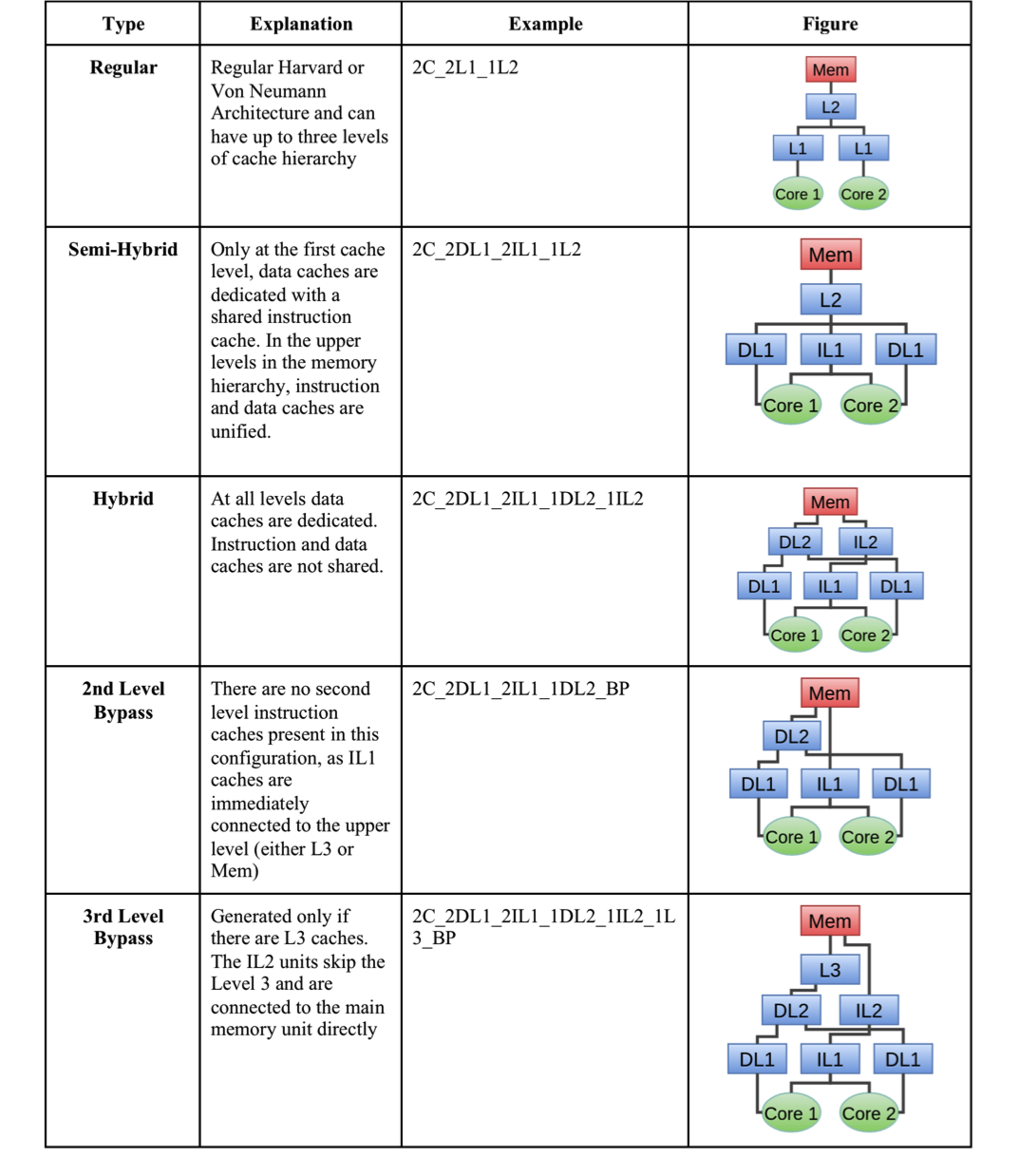} 
\caption{Comparison and explanation of topology methods and their examples}
\label{fig:types}
\end{figure*}

The method types, brief explanations and a two core sample topologies are shown in Figure \ref{fig:types}. These methods are derived from "Topology Creation Rules" \cite{Malazgirt2015}. 
The advantage of decision tree is to extract the maximum available information from the string representation and separate domain information parsing from architecture generation. In Algorithm \ref{alg:topologygen}, we show the architecture generation procedure of Taxim. 

Algorithm \ref{alg:topologygen} starts by parsing the input that is the topology string representation. After string representation is parsed, Taxim tokenizes the cores, instruction/data caches and their levels. Based on this information, the decision tree is traversed and type information of the topology is extracted that is shown in Line 3. Then, the algorithm proceeds by connecting each core to first level data caches. Based on available data caches and instruction caches, each core is either connected to an available cache, or to a cache that has the least amount of connection. After all cores are connected, first level cache connections occur. Each cache level connects to an upper level cache that is free or has the least amount of connection. At the highest cache level, each cache is connected to the main memory. These steps are represented between Line 4 and 5. Based on these steps, the connectivity of the topology can be defined as the following:

Let $nc$ be the number of components at level $i$, $mc$ be the the number of components at level $y$ such that $i<y$, and from Topology Generation Rules \cite{Malazgirt2015}, $nc >= mc$

Components at level i must be connected to components in level y and after each component at level i is connected to an available component at level y, the number of connections $c_{iy}$ created by Algorithm \ref{alg:topologygen} becomes:
\begin{equation}
c_{iy}=\left\lfloor nc/mc\right\rfloor 
\end{equation}
If the number of components is odd, the remaining connections are connected injectively, and  the number of remaining connections $cr_{iy}$ become:
\begin{equation}
cr_{iy}=nc\%mc
\end{equation}
Thus, the total number of connections between level i and y $ct_{iy}$  becomes:
\begin{equation}
ct_{iy} = c_{iy} *mc+cr_{iy}
\end{equation}

After core/cache connections are established, Taxim checks if there are any bypass representation in the given topology representation in Line 6. If there exists any bypassing, Taxim starts to add network switches for forming the bypass structure. The first step is to connect a network switch to the last level memory such as the main memory or last level caches. Then, it starts grouping data caches into pairs. If the number of data caches is odd, the last remaining cache is connected to a pair making which makes it a triple. Then, a new switch is added to an each pair of caches. Similarly, each switch pair is connected to a new switch until all switches are covered. If total number of cache pair switches is odd, it connects to an existing cache switch pair, making it a triple. After data cache switches are created, same steps are applied for instruction caches as well. When all data/instruction cache switches are generated, these switches are connected to the main memory switch. 

 \begin{figure}
\centering
\includegraphics[width=8cm]{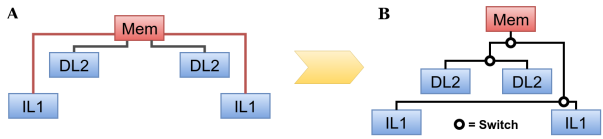} 
\caption{Logical representation (left) and implementation (right) of the network of 5C\_5DL1\_2IL1\_2DL2\_BP}
\label{fig:switchexample}
\end{figure}

Figure \ref{fig:switchexample} visualizes the network creation process explained in Algorithm 1. After Taxim connects DL2 and IL1 caches to main memory in Figure 3A, first a switch is connected to the main memory. Then, DL2 caches are grouped and they are connected to a switch. Next, IL1 caches are also grouped and connected with a switch. The main memory switch is then connected with the cache grouping switches. 

 \begin{figure}
\centering
\includegraphics[width=8cm]{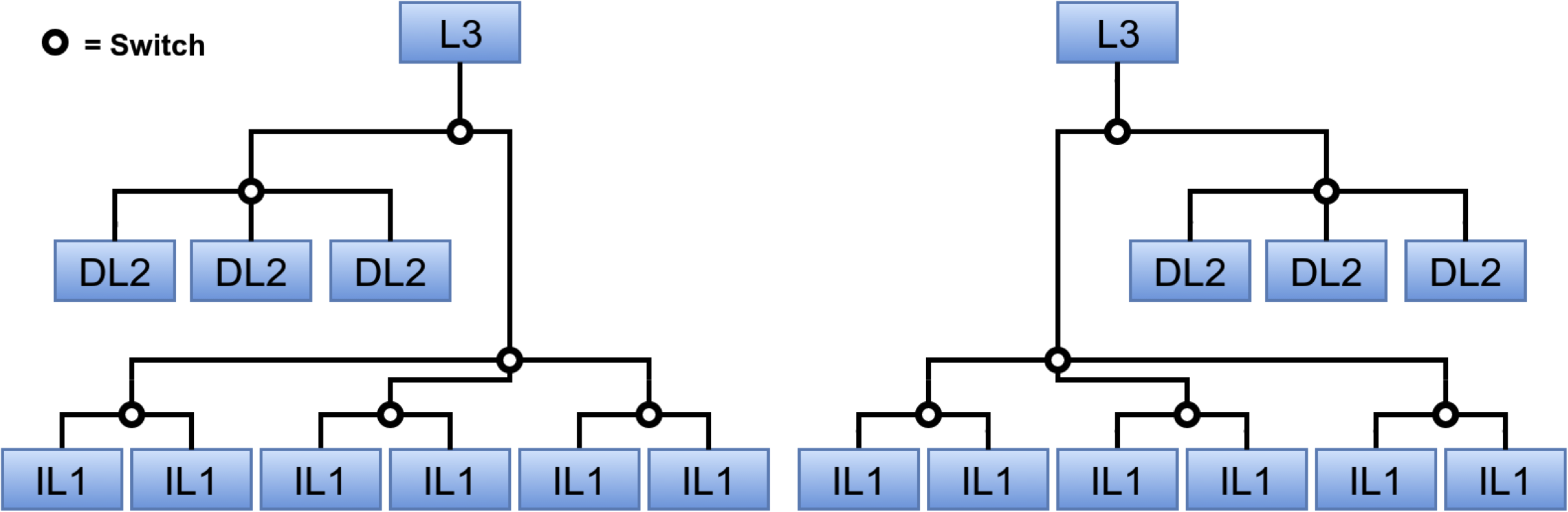} 
\caption{The switch network connections for a bypass structure, there exists no more than three connections from lower level to upper level in the hierarchy}
\label{fig:complexswitch}
\end{figure}

In Figure \ref{fig:complexswitch}, network switch connections of a more complicated topology, 24C\_24DL1\_12IL1\_6DL2\_2L3\_BP (a second level bypass) is shown. First, an L3 cache is connected with a switch. Pairs of IL1 caches are connected with a switch and three IL1 switches are connected to a switch and then connected to the the L3 switch. Three DL2 switches are connected with a switch and then this switch is connected to the L3 switch. These connections are repeated for the second L3 cache, starting with creating a switch for the L3 cache. 

 \begin{figure}[t]
\centering
\includegraphics[width=8cm]{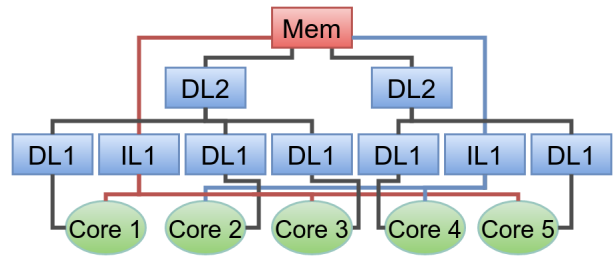} 
\caption{Topology of 5C\_5DL1\_2IL1\_2DL2\_BP Architecture}
\label{fig:fivecoreex}
\end{figure}

As a design example, let us assume that the designer requests the creation of 5C\_5DL1\_2IL1\_2DL2\_BP which is shown in Figure \ref{fig:fivecoreex}. According to the decision tree, first rule is to check whether this topology has a first level data cache (DL1) component. If it is available, the requested topology can be a regular type. Next, the system checks whether a second level data cache (DL2) component also exists and if not, it would be a semi-hybrid memory configuration. In the third rule, it is checked whether the suffix BP has been added to the nomenclature. If true, it is checked whether there is a second level instruction unit. Since in our case there is no second level instruction unit, the system identifies that we have a type "Level 2 Bypass". The bypassing occurs from second level instruction caches to the main memory. 

The topology generation algorithm connects each core in the architecture to a single DL1. Then, each core is connected to a free IL1, however there are only two IL1 in the configuration. Hence, Core 1, 3 and 5 are connected to one IL1 and Core 2 and 4 are connected to the other IL1 as designated by the algorithm. After the cores are connected, first level data caches are connected to second level caches. There are two DL2 in the topology, therefore three DL1 are connected the first DL2 and the other two are connected to the second DL1. Then, there are not any third level caches in the topology, therefore all DL2 are connected to the memory. Since there is no IL2 in the topology, all IL1 are connected to the memory, bypassing second level caches.
Checking component cardinality and connectivity of the generated architecture validates the generated topologies. After given topology string representation is tokenized and cores, data/instruction caches in every cache level are extracted, cardinality of these tokens are compared with the cardinality of the generated cores and caches. For determining the connectivity of the caches and cores in general, the validation method traverses each component in the generated architecture topology and computes the number connections described in Equations (1), (2) and (3). 

As an example, in Figure \ref{fig:fivecoreex}, there exists 5 Cores and 5 DL1 caches, thus each core is connected to a DL1 cache. However, there are 2 IL1 caches, thus after each cache accepts $\left\lfloor 5/2 = 2\right\rfloor $ connections, the remaining $5\%2 = 1$ connection must be connected to the first available cache in the cache list, increasing the number of connections to 3 for one of the IL1 caches. Similarly, at the second cache level, the connectivity between 5 DL1 and 2 DL2 are validated the same way.

\section{Taxim Tool Flow}
\begin{figure}[t]
\centering
\includegraphics[width=8cm]{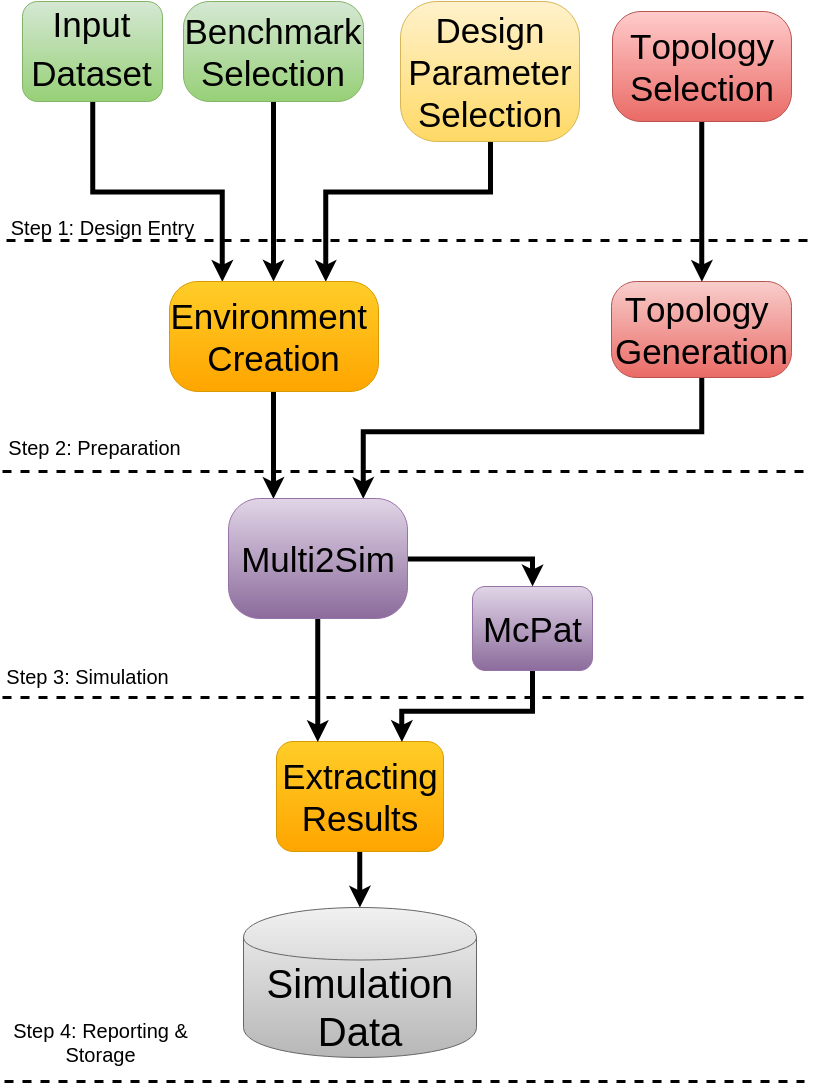} 
\caption{Process flow of Taxim representing all steps from user input until results extraction}
\label{fig:flow}
\end{figure}

In this section we present the process flow of Taxim and detail the important parts. Figure \ref{fig:flow} shows the process flow. The first step is defined as design entry. In this step the designer determines design and simulation specific properties. In the second step, Taxim's stencil scripts read design entry files and create the simulation environment and architectural topologies that are used by Multi2Sim and McPat \cite{Li2009}. Third step is composed of simulation with intelligent control that allows early termination of poor performing design with respect to a user determined design parameter such as latency. The last step includes extraction of simulation results. Results are also stored for later usage for analysis purposes.

The outlined steps are intended to guide the user from designer specifications to aggregated results of numerous parameters that are selected by the designer. As the diagram displays, the designer requirements regulate most of the process flow. The input files are tailored according to chosen topologies and benchmarks, which allow Multi2Sim and McPat \cite{Li2009} to examine their effectiveness. The acquired data is filtered through selected parameters to extract the data for the further evaluation of the designer.
\subsection{Design Entry}
\begin{figure}[t]
\centering
\includegraphics[width=8cm]{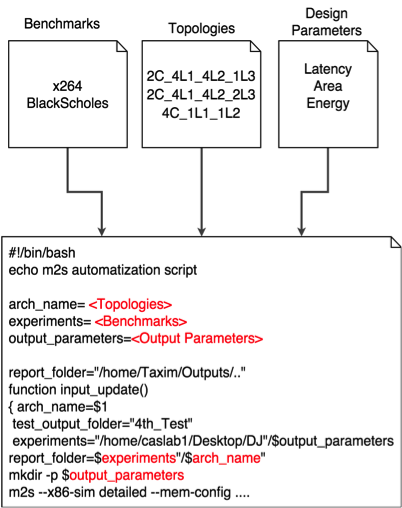} 
\caption{An example environment creation script used by Taxim to allow experimentation in succession}
\label{fig:script}
\end{figure}

In this step, designer enters all the files that Taxim needs for running simulations. As shown in Figure \ref{fig:script}, these are the benchmarks, their input data sets, processor architectures and design parameters that user aims to collect. Benchmarks should be entered in executable form with their required input data sets. Unlike some other platforms in the literature \cite{Binkert2011},\cite{Jia2010}, Taxim accepts plain text format for architecture generation and design parameter selection. Thus, these files can also be edited manually. 

\begin{table}
\begin{tabular}{|c|c|}
\hline 
\textbf{\scriptsize{Component}}{\scriptsize{ }} & \textbf{\scriptsize{Number of Lines}}\tabularnewline
\hline 
\hline 
{\scriptsize{Core }} & {\scriptsize{6}}\tabularnewline
\hline 
{\scriptsize{L1 Cache}} & {\scriptsize{5-6}}\tabularnewline
\hline 
{\scriptsize{L2 and L3 Cache}} & {\scriptsize{6-7}}\tabularnewline
\hline 
{\scriptsize{Main Memory }} & {\scriptsize{5-6}}\tabularnewline
\hline 
{\scriptsize{Cache Geometry Definitions }} & {\scriptsize{5-15}}\tabularnewline
\hline 
{\scriptsize{Cache-Core \& Cache-Cache Connection}} & {\scriptsize{4}}\tabularnewline
\hline 
{\scriptsize{Single Ring Bus Network Element}} & {\scriptsize{14}}\tabularnewline
\hline 
\end{tabular}
\caption{Lines of code required per component} \label{tbl:comploc}
\end{table}

\subsection{Simulation Preparation}
In order to support a broad range of systems, Taxim provides automation bash scripts that can be executed on operating systems that support bash. Taxim provides input and output locations which Multi2Sim \cite{Ubal2007} and McPat \cite{Li2009} simulators read inputs and write output values. Automation scripts are initially in stencils that are found in the environment. They are processed after the designer manually completes them. As an example, in Figure 8, we present a stencil which relates the scripts to the Design Entry files that are shown as Benchmarks, Topologies and Design Parameters. In the Design Entry step, the designer enters the locations of the mentioned design entry files, thus Taxim populates the stencil scripts. This design also allows the designer to conduct experiments in parallel by calling multiple instances in the terminal or adding related scripts in the same file.

\subsection{Simulation and Intelligent Control}
In this step, Taxim starts Multi2Sim \cite{Ubal2007} architectural simulation. Following the architectural simulation, power, area and timing simulations require McPAT \cite{Li2009} simulator. In order to customize McPat \cite{Li2009}, Multi2Sim simulation reports are extracted. McPat \cite{Li2009} requires simulation output that represents all the components that are present in the architecture. These are datapath details of each core such as the number of instructions in the pipeline stages, cache and memory communication details such as cache miss/hit ratio, and interconnection network details such as sent and received packets etc. The output of McPat \cite{Li2009} simulation are located in designer designated locations and the results are reported together with performance outputs. The designer does not need to do any additional work in order to extract McPat \cite{Li2009} output. The output data extraction and reporting is explained in Section 4.5.  

Taxim introduces two simple methods for early termination of simulation: 
\begin{itemize}
\item While setting up the environment for multiple simulations, the designer might enter some parameters wrong. Hence, the simulations have to be aborted. Taxim prevents designers to simulate erroneous tests. When a very large number of experiments are configured with many benchmarks, input data sets and topologies, the designer can opt for first testing the benchmarks, their input and the topologies by running a functional simulation that is a lot faster than cycle accurate simulation. If there exists no errors during functional simulation, then the cycle accurate simulation starts. 
\item The designer can provide a design parameter and a satisfaction condition. If the condition is not met after a predetermined simulation time which is chosen by the designer, Taxim drops the simulation and proceeds with the next topology in the simulation queue. For example, user can put a condition on Latency metric. If the desired Latency is not met after a certain amount of simulation time, Taxim stops simulating the design under test. 
\end{itemize}

\begin{table*}[t]
\centering
\begin{tabular}{|c|c|c|c|}
\hline 
\textbf{\scriptsize{Component}}{\scriptsize{ }} & \textbf{\scriptsize{Basic Topology}} & \textbf{\scriptsize{LOC}} & \textbf{\scriptsize{Generation Time (ms)}}\tabularnewline
\hline 
\hline 
{\scriptsize{Regular }} & {\scriptsize{2C\_2L1\_2L2\_1L3 }} & {\scriptsize{76}} & {\scriptsize{9.09 }}\tabularnewline
\hline 
{\scriptsize{Semi - Hybrid}} & {\scriptsize{4C\_4DL1\_2IL1\_1L2 }} & {\scriptsize{83}} & {\scriptsize{8.9}}\tabularnewline
\hline 
{\scriptsize{Hybrid}} & {\scriptsize{3C\_3DL1\_3IL1\_3DL2\_1IL2\_1L3 }} & {\scriptsize{122}} & {\scriptsize{8.74}}\tabularnewline
\hline 
{\scriptsize{2nd Level Bypass }} & {\scriptsize{2C\_2DL1\_2IL1\_1DL2\_1L3\_BP }} & {\scriptsize{130}} & {\scriptsize{9.24}}\tabularnewline
\hline 
{\scriptsize{3rd Level Bypass }} & {\scriptsize{4C\_4DL1\_2IL1\_2DL2\_2IL2\_1L3\_BP }} & {\scriptsize{182}} & {\scriptsize{9.28}}\tabularnewline
\hline 
\hline 
\textbf{\scriptsize{Component}}{\scriptsize{ }} & \textbf{\scriptsize{Complex Topology}} & \textbf{\scriptsize{LOC}} & \textbf{\scriptsize{Generation Time (ms)}}\tabularnewline
\hline 
{\scriptsize{Regular }} & {\scriptsize{13C\_9L1\_5L2\_3L3 }} & {\scriptsize{227}} & {\scriptsize{8.64}}\tabularnewline
\hline 
{\scriptsize{Semi - Hybrid}} & {\scriptsize{18C\_9DL1\_6IL1\_3L2 }} & {\scriptsize{232}} & {\scriptsize{8.65}}\tabularnewline
\hline 
{\scriptsize{Hybrid}} & {\scriptsize{17C\_11DL1\_8IL1\_5DL2\_3IL2\_2L3 }} & {\scriptsize{321}} & {\scriptsize{9.09 }}\tabularnewline
\hline 
{\scriptsize{2nd Level Bypass }} & {\scriptsize{32C\_23DL1\_17IL1\_12DL2\_4L3\_BP}} & {\scriptsize{1105}} & {\scriptsize{9.39 }}\tabularnewline
\hline 
{\footnotesize{3rd Level Bypass }} & {\footnotesize{37C\_28DL1\_19IL1\_13DL2\_8IL2\_5L3\_BP}} & {\footnotesize{979}} & {\footnotesize{9.28 }}\tabularnewline
\hline 
\end{tabular}
\caption{Examples of for each topology, followed by the amount of lines and
generation times } \label{tbl:loc}
\end{table*}

\subsection{Reporting and Storage}
Taxim extracts necessary output parameters from the simulation output files. Taxim recognizes all design parameters that Multi2Sim \cite{Ubal2007} and McPat \cite{Li2009} provide. Currently, Taxim supports Comma Separated Values (csv) and plain text file outputs to store these design parameters. The output locations and designated design parameters are configured by the designer in Step 1 that is shown in Figure \ref{fig:flow}. The generated results can be used for exploration \cite{Malazgirt2015}, augmented to DSE tools \cite{Zaccaria2010} or stored in databases. 

\section{Experiments}
In this section, we present how our automated tool improves the overall simulation process that can be used as a standalone or part of a DSE tool. We have used Taxim in exploring various symmetric multiprocessing systems \cite{Malazgirt2015}.  During our experiments in \cite{Malazgirt2015}, 500 design points have been considered for simulation. These simulations involved around 100 different topologies and 5 different test cases from PARSEC \cite{Bienia2008} and MiBench \cite{Guthaus2001}  benchmarks.  

\subsection{Topology Generation Metrics}
The lines of simulation codes generated by the topology generation algorithm is proportional to the the topology. As an example, a complex topology such as 20C\_10DL1\_4IL1\_5DL2\_2L3\_BP requires much more lines of code (LOC) for connectivity than a simple 2C\_2DL1\_2IL1\_2DL2\_2IL2\_1L3 architecture. Taxim applies Algorithm \ref{alg:topologygen} in order to build Multi2Sim \cite{Ubal2007} topology configuration file of the 20-core architecture. Thus, all the components and connectivity structures are defined. In Table \ref{tbl:comploc}, we present the number of lines that is required to describe components and connections between them. Hence, increasing number of components and bypass topologies take much more lines of code to represent. According to Table \ref{tbl:loc}, the sample 20-Core design takes 443 LOC. However, for 2-core architecture, it is only 106 LOC. Thus, the automation Taxim provides will be more visible and helpful for future non-regular complex topologies.

To further illustrate this point, we have created a basic and very complicated model of each topology method using Taxim. Creating a topology list took less than five minutes and all the topologies were created within 70 milliseconds. Manual construction would take five to thirty minutes for each basic topology and an indefinite time for very complicated topologies.

Comparing the LOC required by 13C\_9L1\_5L2\_3L3 and 18C\_9DL1\_6IL1\_3L2, which is only a 5 line difference, it is understandable that the complexity between topologies is not entirely dependent on the amount of cores, but the amount of connections that have to be defined for further modules. Hence, 18C\_6L1\_3L2 is very easy to code, but 6C\_4DL1\_3IL1\_2L2 is a bigger challenge due to varying input and output streams between caches. The same is true for complicated networks too, as 33C\_23DL1\_17IL1\_12DL2\_4L3\_BP has a higher line count than 37C\_28DL1\_19IL1\_13DL2\_8IL2\_5L3\_BP. This is mainly due to their network files, since 33C\_23DL1\_17IL1\_12DL2\_4L3\_BP  17 IL1 caches bypass second and third level and connect to memory, whereas in  37C\_28DL1\_19IL1\_13DL2\_8IL2\_5L3\_BP  8 IL2 bypass third level cache and connect to the memory. According to Table III, network switches that are used in bypass topologies consume the most lines of code, therefore 33C\_23DL1\_17IL1\_12DL2\_4L3\_BP consumes more lines than 37C\_28DL1\_19IL1\_13DL2\_8IL2\_5L3\_BP. 

Table \ref{tbl:loc} also illustrates that, managing even simple topologies requires diligence for many connections, and creating topologies with higher complexity manually is prone to errors. Hence, by automating the creation and validation, we can eliminate mistakes, optimize the topology creation process and interconnections of topologies themselves.



\textbf{Intelligent Simulation Control}: The intelligent simulation control capabilities of Taxim that are explained in Section 4.3, have saved considerable amount simulation time in our experiments. As an example, Vips \cite{Bienia2008} benchmark on 4-core bypass and hybrid architectures takes 15 hours on average to complete in our simulation test bed. Using early termination method, we have opted to simulate Vips benchmarks on topologies which yield instructions per cycle (IPC) value greater or equal to 1.0. After thirty minutes of initial simulation,  topologies that do not meet the IPC criteria have been eliminated. Thus, 10 topologies out of 64 had IPC less than 1.0. Remaining 54 topologies have been simulated with cycle accurate simulation. This elimination has reduced the overall simulation time by 12.2

Another aspect of Taxim's intelligent control is to prevent wasting simulation time due to erroneous design entry by the designer. This is overcome by first running functional simulation on each topology before the cycle-accurate simulation \cite{Ubal2007}. The functional simulation emulates underlying architecture and takes significantly less amount of time compared to detailed simulation time. In our testbed, x264 detailed simulation on a 4-core architecture takes 12 hours on average to complete whereas the functional simulation takes 6 minutes. Assuming that 60 simulations will be carried out sequentially, and 10 simulations are faulty due to erroneous design entry by the designer. If pre-check functional simulation had not been opted, 120 hours of simulation time would be wasted. However, by opting pre-check, the designer can fix the faults after functional simulations. Thus, in this experimental scenario, intelligent pre-check simulation saved 114 hours overall simulation time.
\section{Conclusion}
In this paper, we present Taxim, a toolchain that automates architectural, power, area and timing simulation using Multi2Sim \cite{Ubal2007} and McPat \cite{Li2009} tools. Taxim provides significant speed up for creating processor architectures and output generation compared to manual methods.  
Decision tree based automatic topology generation allows generating complex multi core architectures from simple text representations. Intelligent control mechanism embedded in Taxim prevents wasting simulation time. Taxim source code is available for use at \textit{https://github.com/bouncaslab/TaXim/}. In our experiments with taxim,  we have shown that complex multiprocessing architectures can get more than a thousand lines in Multi2Sim's representation format, which is very difficult and slow to generate manually. However, Taxim can generate complex architectures in milliseconds. Alongside automatic processor generation, intelligent control prevents designers to save valuable simulation time due to human errors.




\section{Acknowledgments}
Funding from the Turkish Ministry of Development under the TAM Project, number 2007K120610 and Bogazici University Scientific Projects number 7060 was received.

%
\bibliographystyle{abbrv}

\balance

\end{document}